%% file: main.tex
\documentclass[12pt]{llncs}
\usepackage[boxed]{algorithm}
\usepackage{algorithmic}
\usepackage{floatflt}
\usepackage{amsfonts}
\usepackage{amssymb}
\usepackage{amstext}
\usepackage{complexity}
\usepackage[nofancy,notoday]{rcsinfo}
\input{macros.tex}

\begin{document}

\bibliographystyle{plain}

\begin{titlepage}
\pagestyle{plain}

\begin{center}
\vspace*{4cm}
{\Large\bf The Complexity of Computing Optimal Assignments of
Generalized Propositional Formulae}
\\[2cm]

{\large\it Stef{f}en Reith \quad and \quad Heribert Vollmer}
\\[.5cm]

 Lehrstuhl f\"ur Theoretische Informatik\\
 Universit\"at W\"urzburg\\
 Am Exerzierplatz 3\\
 D-97072 W\"urzburg, Germany\\
 \email{\symbol{"5B}streit,vollmer\symbol{"5D}@informatik.uni-wuerzburg.de}
\\[2cm]

{\normalsize\rm Running head: The Complexity of Computing Optimal Assignments}

\end{center}

\end{titlepage}

\rcsInfo $Id: main.tex,v 1.17 1998/03/06 13:02:46 vollmer Exp $

%\lhead{\tiny Rev.: \rcsInfoRevision\ Date: \rcsInfoDate\ Time: \rcsInfoTime}
%\lhead{}
%\chead{}
%\rhead{}
%\lfoot{}\cfoot{\thepage}\rfoot{}
%\lfoot{}
%\cfoot[\thepage]{\thepage}
%\rfoot{}
%\setlength{\headrulewidth}{0pt}
%\setlength{\headrulewidth}{0.8pt}
%\setlength{\footrulewidth}{0pt}
%\setlength{\plainheadrulewidth}{0pt}
%\setlength{\plainfootrulewidth}{0pt}
%\pagestyle{fancy}

%\parindent0.5em
%\parskip1ex plus.5ex minus.2ex

%
%* Abstract
%\input{abstract.tex}
%

\vspace*{2cm}

\centerline{\large\bf R\'esum\'e}

Nous \'etudions le probl\`eme du calcul de l'affectation bool\'eenne
minimale (ou maximale, selon l'ordre lexicographique) satisfaisant une
expression bool\'eenne, en fonction de restrictions impos\'ees \`a la
classe des expressions consid\'er\'ees.  Il appert que pour chaque classe
envisag\'ee, le probl\`eme est soit r\'esoluble en temps polynomial,
soit complet pour $\OptP$.  Nous examinons \'egalement le probl\`eme
de d\'ecider si la plus grande variable prends la valeur $1$ selon
l'affectation optimale.  Nos montrons que ce probl\`eme est soit dans
$\P$, soit complet pour $\P^{\NP}$.

\bigskip
 
\noindent\emph{Mots-cl\'es:} Complexit\'e du calcul,
probl\`emes de satisfaisabilit\'e,
optimisation.

\vspace*{2cm}

\centerline{\large\bf Abstract}

We consider the problems of finding the lexicographically
minimal (or maximal) satisfying assignment of propositional
formulae for different restricted
formula classes. It turns out that for each class from our
framework, the above problem is either polynomial time solvable
or complete for $\OptP$. We also consider the problem of
deciding if in the optimal assignment the largest variable gets
value $1$. We show that this problem is either in $\P$ or
$\P^{\NP}$ complete.
\bigskip


\newpage

%* Introduction 
\input{intro.tex}

%* Preliminaries
\input{pre.tex}

%* Some easy observations
\input{easy.tex}

%* Main results
\input{mres.tex}

%* FP^NP
\input{fpnp.tex}

%* Conclusion
\input{concl.tex}

%* Literatur
\bibliography{main}

%\layout

\end{document}

%% file: macros.tex
\rcsInfo $Id: macros.tex,v 1.13 1998/03/06 13:02:45 vollmer Exp $
\newcommand{\lms}{\ensuremath{\mathit{Lex\-Min}\-\SAT}\xspace}
\newcommand{\LMSN}[1]{\ensuremath{\mathit{Lex\-Min}\-\SAT_{NC}(#1)}\xspace}
\newcommand{\LMS}[1]{\ensuremath{\mathit{Lex\-Min}\-\SAT_{C}(#1)}\xspace}

\newcommand{\LMT}[0]{\ensuremath{\mathit{Lex\-Min}\-\TSAT}\xspace}
\newcommand{\las}{\ensuremath{\mathit{Lex\-Max}\-\SAT}\xspace}
\newcommand{\LASN}[1]{\ensuremath{\mathit{Lex\-Max}\-\SAT_{NC}(#1)}\xspace}
\newcommand{\LAS}[1]{\ensuremath{\mathit{Lex\-Max}\-\SAT_{C}(#1)}\xspace}
\newcommand{\omats}{\ensuremath{\mathit{Odd\-Min}\-\TSAT}\xspace}
\newcommand{\omas}{\ensuremath{\mathit{Odd\-Min}\-\SAT}\xspace}
\newcommand{\OMAS}[1]{\ensuremath{\mathit{Odd\-Min}\-\SAT_{C}(#1)}\xspace}
\newcommand{\OMASN}[1]{\ensuremath{\mathit{Odd\-Min}\-\SAT_{NC}(#1)}\xspace}

\newcommand{\oaas}{\ensuremath{\mathit{Odd\-Max}\-\SAT}\xspace}
\newcommand{\OAAS}[1]{\ensuremath{\mathit{Odd\-Max}\-\SAT_{C}(#1)}\xspace}
\newcommand{\OAASN}[1]{\ensuremath{\mathit{Odd\-Max}\-\SAT_{NC}(#1)}\xspace}
\newcommand{\zval}[0]{\ensuremath{$0$\mbox{-valid}}\xspace}
\newcommand{\nval}[0]{\ensuremath{$1$\mbox{-valid}}\xspace}
\newcommand{\var}[1]{\ensuremath{\mathit{Var}(#1)}\xspace}
\newcommand{\gen}[1]{\ensuremath{\mathit{Gen_{C}}(#1)}\xspace}
\newcommand{\rep}[1]{\ensuremath{\mathit{Rep_{C}}(#1)}\xspace}
\newcommand{\genn}[1]{\ensuremath{\mathit{Gen_{NC}}(#1)}\xspace}
\newcommand{\repn}[1]{\ensuremath{\mathit{Rep_{NC}}(#1)}\xspace}
\newcommand{\SATSNN}[1]{\ensuremath{\SAT(#1)}\xspace}
\newcommand{\SATS}[1]{\ensuremath{\SAT_{C}(#1)}\xspace}
\newcommand{\SATSN}[1]{\ensuremath{\SAT_{NC}(#1)}\xspace}
\newcommand{\SATSS}[1]{\ensuremath{\SAT^{*}(#1)}\xspace}

\newlength\problength
\newcommand\prob[3]{%
\begin{list}{}{\labelwidth\problength \labelsep.5em \rightmargin0em
\leftmargin\problength \advance\leftmargin by2.3em
\parsep0ex \itemsep.2ex plus.1ex}
\item[{\textsc{Problem:}  \hfill}] #1
\item[{\textsc{Instance:} \hfill}] #2
\item[{\textsc{Output:}  \hfill}] #3
\end{list}
}

%% file: intro.tex
%-*-LaTeX-*-
%
\rcsInfo $Id: intro.tex,v 1.7 1998/03/06 13:02:44 vollmer Exp $
%
%*Introduction
\section{Introduction}

In 1978 Thomas J.~Schaefer proved a remarkable result. 
He examined satisfiability
of propositional formulae for certain syntactically restricted formula
classes. Each such class is given by a set $S$ of boolean relations allowed
when constructing formulae. 
An \emph{$S$-formula} is a conjunction of clauses, where each clause
consists out of a relation from $S$ applied to some propositional variables.
$\SATSNN{S}$ now is the problem to decide for a given $S$-formula if it is
satisfiable.
Schaefer showed that depending on $S$ the problem $\SATSNN{S}$ is either (1)
efficiently (i.\,e.~polynomial time) computable or 
(2) $\NP$-complete; and he gave
a simple criterion that, given some $S$, allows to determine
whether (1) or (2) holds.
Since (depending on $S$) the complexity of $\SATSNN{S}$ is either easy or hard 
(and there is nothing in between), Schaefer called this a 
``dichotomy theorem for satisfiability.''

In the last few years his result regained interest among complexity theorists.
In 1995 Nadia Creignou examined the problem of determining the maximal
number of clauses of a given $S$-formula that can be satisfied simultaneously.
Interestingly she also obtained a dichotomy theorem:
She proved that this problem is either polynomial-time solvable 
or $\MaxSNP$-complete, depending on properties of $S$ \cite{cre95}. 
(In 1997 the approximability of this
problem and the corresponding minimization problem was examined in 
\cite{khsuwi97,khsutr97}, leading to a number of deep results.) 
The complexity of counting problems and enumeration problems based on 
satisfiability of $S$-formulae was examined in \cite{crhe96,CrHe97}.

The problem of maximizing (or minimizing) the number of clauses satisfied
in (unrestricted) propositional formula is complete for the class $\MaxSNP$ 
(or $\MinSNP$). These classes, 
introduced in 1988 by Papadimitriou and Yannakakis \cite{PY88}
(see also \cite[pp.~311ff]{pap94}), are of immense importance in the theory
of approximability of hard optimization problems.
Of equal importance however is the class $\OptP$, introduced by Krentel
in 1988 \cite{kre88}. While $\MaxSNP$ and $\MinSNP$ are defined logically
making use of Fagin's characterization of $\NP$ \cite{fag74}, the
class $\OptP$ is defined using Turing machines.
$\OptP$ is a superclass of $\MaxSNP$ and $\MinSNP$. 
The canonical complete problems for $\OptP$ are the problems
$\las$ and $\lms$ of determining the
lexicographically maximal (or minimal) satisfying assignment of a given
(unrestricted) propositional formula.

In this paper we examine $\las$ and $\lms$ for classes of $S$-formulae.
We show that both problems are either polynomial-time
solvable or $\OptP$ complete, depending on properties of $S$. That is, we
prove a dichotomy theorem for the \las (and \lms) 
problem. 
Comparing our results with those of Schaefer we gain insight in the 
connection between the complexity of a decision problem and the 
corresponding optimization problem. We show for example that
if constants are allowed in $S$ formulae, then the problem of deciding
satisfiability is $\NP$-complete if and only if the problem of finding
the smallest assignment is $\OptP$-complete. 
(In the case that constants are forbidden, an analogous result does not
hold unless $\P=\NP$.)

Generally the connection between decision problems and optimization
problems is open. It can very well be that an optimization problem
is hard (complete) though the decision problem is trivial. Here we
show that in the case that constants are allowed, this cannot happen:
a decision problem is hard if and only if the corresponding optimization
problem is hard. In contrast to this, if constants are forbidden then
we completely identify those cases where the optimization problem is
hard and the decision problem is easy. We hope that these results
help to better understand the connection between the complexity of
decision problems and optimization problems.

From an $\OptP$-complete optimization problem one can sometimes obtain a 
decision problem that is complete for $\P^{\NP}$. In our case this is the
$\omas$ (or $\oaas$) problem, for an exact definition refer to 
Sect.~\ref{fpnp}.
We prove that this problem is either polynomial-time
solvable or complete for $\P^{\NP}$; that is we again get a dichotomy
theorem.

%% file: pre.tex
%-*-LaTeX-*-
%
\rcsInfo $Id: pre.tex,v 1.23 1998/03/06 13:02:48 vollmer Exp $
%
%*Preliminaries

\section{Preliminaries}

\label{pre}

Any subset $R \subseteq \{0,1\}^k$ is called a \emph{\kary{k} boolean relation}
(\emph{\kary{k} logical relation}). The integer $k$ is called the \emph{rank}
of $R$. If $k$ is not needed or is clear from the context we use
\emph{boolean relation} (\emph{logical relation}) for short. Since we
need symbols representing boolean relations in the formulae we construct, 
we always use lowercase
letters for relation symbols and uppercase letters for the relation
itself. So the relation symbol $r$ represents the relation $R$.

We will consider different types of relations, following the terminology
of Schaefer \cite{Sch87}.

\begin{enumerate}
\item
The boolean relation $R$ is \emph{\zval} 
(\emph{\nval}, resp.) \gdw $(0, \ldots , 0) \in R$ ($(1, \ldots , 1)
\in R$, resp.).

\item
The boolean relation $R$ is \emph{Horn} (\emph{anti-Horn}, resp.) \gdw
$R$ is logically equivalent to a CNF formula having at most one unnegated 
(negated, resp.) variable in any conjunct.

\item
A boolean relation $R$ is \emph{bijunctive} \gdw it is logically
equivalent to a CNF formula having one or two variables in each
conjunct.

\item
The boolean relation $R$ is \emph{affine} \gdw it is logically
equivalent to a system of linear equations over the finite 
field $\Z_2$. This means that any tuple $(v_1, \ldots ,v_k) \in R$ is a
solution of a system of formulae of the form 
$x_1 \oplus x_2 \oplus \cdots \oplus x_n = 0$ or 
$x_1 \oplus x_2 \oplus \cdots \oplus x_n = 1$.
\end{enumerate}

Now let $S = \{R_1, \ldots ,R_n \}$ be a set of boolean
relations. In the rest of this paper we will always assume that such
$S$ are nonempty and finite.
$S$ is called \zval (\nval, Horn, anti-Horn, affine,
bijunctive, resp.) \gdw every relation $R_i \in S$ is \zval (\nval, Horn, 
anti-Horn, affine, bijunctive, resp.).

$S$ formulae will now be propositional formulae consisting of clauses
built by using relations from $S$ applied to arbitrary variables.
Formally, let $S = \{R_1, R_2, \ldots , R_n \}$ be a set of logical
relations and $V$ be a set of variables. We will always assume
an ordering on $V$. 
An \emph{$S$-formula} $\Phi$ (over $V$) is a finite conjunction of 
clauses $\Phi = C_1 \wedge \ldots \wedge C_k$, where each $C_i$ is of the
form $r(\enu{x}{1}{k})$, $R \in S$, $r$ is the symbol
representing $R$, $k$ is the rank of $R$, and $\enu{x}{1}{k} \in V$.
If some variables of an $S$-formula $\Phi$ are replaced by the 
constants $0$ or $1$ then this new formula $\Phi'$ is called 
\emph{$S$-formula with constants}. By $\var{\Phi} \subseteq V$ we denote
the subset of those variables actually used in $\Phi$.

The satisfiability problem for $S$-formulae ($S$-formulae with
constants, resp.) is denoted by $\SATSN{S}$ ($\SATS{S}$, resp.).

By $\subst{\Phi}{x}{\smash{y}\vphantom{1}}$ 
%By $\subst{\Phi}{x}{y}$
we denote the formula created by simultaneously
replacing each occurrence of $x$ in $\Phi$ by $y$, where $x,y$
are either variables or a constants. 
%If $X$ is a set of variables, then $\subst{\Phi}{X}{y}$ denotes 
%the formula $(\subst{(\subst{(\subst{\Phi}{x_1}{y})}{x_2}{y}) 
%\ldots}{x_n}{y})$ where $X = \{\enu{x}{1}{n} \}$.
Now we define the set of \emph{existentially quantified $S$-formulae
with constants}, again following Schaefer. 
Let $\gen{S}$ the smallest set of formulae
having the following closure properties: 
For any $k \in \N$ and any $k$-ary relation $R \in S$ 
where $\enu{x}{1}{k} \in V$, the formula $r(\enu{x}{1}{k})$ is in $\gen{S}$. 
Now let $\Phi$ and $\Psi$ be in $\gen{S}$, $x,y\in V$, 
then $\Phi \wedge \Psi$, 
$\subst{\Phi}{x}{\smash{y}\vphantom{1}}$, $\subst{\Phi}{x}{1}$, 
$\subst{\Phi}{x}{0}$ and $(\exists x) \Phi$ are in $\gen{S}$, 
for $x,y \in V$. Define $\genn{S} \eqd \{ \Phi | \Phi \in \gen{S} 
\text{ and $\Phi$ has no constants} \}$.
For $\Phi\in\gen{S}$ let $\var{\Phi}$ be the set of variables with
free occurrences in $\Phi$.

Let $\Phi$ be an $S$-formula with $k$ variables. If 
$\var{\Phi}=\{\enu{x}{1}{k}\}$,
$x_1<\cdots<x_k$ (recall that $V$ is ordered), 
then an assignment $I\colon\var{\Phi}\rightarrow\{0,1\}$ 
where $I(x_i)=a_i$ will also be denoted by $(\enu{a}{1}{k})$.
The ordering on variables induces an ordering on assignments as follows:
$(\enu{a}{1}{k})<(\enu{b}{1}{k})$ if and only if there is 
an $i\leq k$ such that for all $j < i$ we have $a_j=b_j$ and $a_i<b_i$.
We refer to this ordering as the \emph{lexicographical ordering}.
That an assignment $(\enu{a}{1}{k}) 
\in \{0,1\}^k$ satisfies $\Phi$ will be denoted 
by $(\enu{a}{1}{k}) \models \Phi$.
We write $(\enu{a}{1}{k}) \minmodels \Phi$ ($(\enu{a}{1}{k}) \maxmodels 
\Phi$, resp.) iff $(\enu{a}{1}{k}) \models \Phi$ and there exists no 
lexicographically smaller (larger, resp.) $(\enu{a'}{1}{k}) \in \{0,1\}^k$ 
such that $(\enu{a'}{1}{k}) \models \Phi$.
If $I\colon\{x_1,x_2,\dots\}\rightarrow\{0,1\}$ is an arbitrary assignment,
$y$ is variable and $a\in\{0,1\}$, then $I\cup\{y:=a\}$ denotes the
assignment $I'$ defined by $I'(y)=a$ and $I'(x)=I(x)$ for all $x\neq y$.

Let $\relof{\Phi} \eqd \{ (\enu{a}{1}{k}) \in \{0,1\}^k
| \var{\Phi} = \{\enu{x}{1}{k}\} \text{ and } (\enu{a}{1}{k}) \models \Phi \}$
be the logical relation defined by $\Phi$, and let
\begin{displaymath}
\begin{array}{rcl}
\rep{S} &\eqd& \{ \relof{\Phi} | \Phi \in \gen{S} \}\\
\repn{S} &\eqd& \{ \relof{\Phi} | \Phi \in \genn{S} \}
\end{array}
\end{displaymath}
%
% Ende der Definitionen
%

The following results proved by Schaefer will be needed in this paper.

\begin{proposition}[\cite{Sch87}, Theorem 3.0]
\label{RepIsAll}
Let $S$ be a set of logical relations. If $S$ is Horn, anti-Horn,
affine or bijunctive, then \rep{S} satisfies the same condition. 
Otherwise, \rep{S} is the set of all logical relations. 
\end{proposition}

\begin{proposition}[\cite{Sch87}, Lemma 4.3]
\label{StructOfRepS}
Let $S$ be a set of logical relations. Then at least one of the
following four statements holds:
\begin{enumerate}
\def\labelenumi{(\arabic{enumi})}
\item $S$ is \zval
\item $S$ is \nval
\item $[ x ], [ \neg x ] \in \repn{S}$
\item $[ x \not\equiv y ] \in \repn{S}$
\end{enumerate}
%Furthermore, if statement \textit{(iii)} does not hold, but \textit{(iv)} 
%holds, then every relation $R_i \in S$ is \emph{complementative}. This
%means that if $(\enu{x}{1}{n}) \in R_i$ then also $(\enu{1-x}{1}{n}) \in R_i$.
\end{proposition}

Schaefer's main result, a dichotomy theorem for satisfiability of
propositional formulae, can be stated as follows:

\begin{proposition}[Dichotomy Theorem for Satisfiability with Constants]
%\cite{Sch87}]%, Lemma 4.1]
\label{SATDich1}
Let $S$ be a set of logical relations. If $S$ is Horn, anti-Horn, 
affine or bijunctive, then $\SATS{S}$ is polynomial-time
decidable. Otherwise $\SATS{S}$ is $\NP$-complete.%$\log$-complete in $\NP$.
\end{proposition}

\begin{proposition}[Dichotomy Theorem for Satisfiability]
%\cite{Sch87}]%, Theorem 2.1]
\label{SATDich2}
Let $S$ be a set of logical relations. If $S$ is \zval, \nval,
Horn, anti-Horn, affine or bijunctive, then $\SATSN{S}$ is polynomial-time
decidable. Otherwise $\SATSN{S}$ is $\NP$-complete.%$\log$-complete in $\NP$.
\end{proposition}

By \SATSS{S} we denote the problem to decide whether there exists
a satisfying assignment for an $S$-formula which is different from 
$(0,0, \ldots ,0)$ and $(1,1, \ldots ,1)$.
The following proposition is from Creignou and H\'ebrard
\cite{CrHe97}.

\begin{proposition}
\label{SATSSNPC}
Let $S$ be a set of logical relations.
If $S$ is not Horn, anti-Horn, affine and bijunctive, then \SATSS{S} is 
\NP-complete.
\end{proposition}

%% file: easy.tex
%-*-LaTeX-*-
%
\rcsInfo $Id: easy.tex,v 1.24 1998/03/06 13:02:42 vollmer Exp $
%
%*Min and Max Problems

\section{Maximization and Minimization Problems}
\label{easy}

The study of optimization problems in computational complexity theory
started with the work of Krentel \cite{kre88,kre92}. He defined the 
class $\OptP$  and an oracle hierarchy built on this class using so 
called metric Turing machines. We do not need this machine model here;
therefore we proceed by defining the classes relevant in our context 
using a characterization given in \cite{vowa95}.

We fix the alphabet $\Sigma=\{0,1\}$. 
Let $\FP$ denote the class of all functions $f\colon\Sigma^*\rightarrow
\Sigma^*$ computable deterministically in polynomial time. 
Using one of the well-known bijections between $\Sigma^*$ and the set
of natural numbers (e.g.~dyadic encoding) we may also think of $\FP$
(and the other classes of functions defined below) as a class of
number-theoretic functions.
Say that a function $h$ belongs to the class
$\MinP$ if there is a function $f \in \FP$ and a polynomial $p$ such 
that for all $x$,

\begin{displaymath}
h(x) = \min_{|y| \le p(|x|)}f(x,y).
\end{displaymath}

The class $\MaxP$ is defined by taking the maximum of these values.
Finally, let $\OptP = \MinP\cup\MaxP$.

Krentel considered the following reducibility in connection with these
classes: 
A function $f$ is \emph{metric reducible} to $h$ ($f \redm h$) if there 
exist two functions $g_1,g_2 \in FP$ such that for all $x$:
$$f(x) = g_1(h(g_2(x)),x).$$ 
As a side remark let us mention that the closure of all three classes
$\MinP$, $\MaxP$, and $\OptP$ under metric reductions coincides with the
class $\FP^{\NP}$; which means that showing completeness of a problem for
$\MinP$ generally implies hardness of the same problem for $\MaxP$ and
completeness for $\OptP$, see \cite{kre88,vowa95,vol94b}.

Krentel gave in \cite{kre88} a number of problems complete for $\OptP$
under metric reducibility. The for us most important complete problem
for $\OptP$ is the problem of finding the lexicographically minimal
satisfying assignment of a given formula. 

\prob{\lms}
{a propositional formula $\Phi$}
{the lexicographically smallest satisfying assignment of $\Phi$}

The problem \las is defined analogously.

\begin{proposition}[\cite{kre88}]\label{krentelcomplete}
\lms and \las are complete for $\OptP$ under metric reductions.
\end{proposition}

One of the main points of this paper is to answer the question
for what syntactically restricted classes of formulae (given by a set
$S$ of boolean relations) the above proposition 
remains valid. For this, we will consider the following problems:

\prob{Lexicographically Minimal \SAT  ($\LMSN{S}$)}
{An $S$-formula $\Phi$}
{The lexicographically smallest satisfying  assignment of $\Phi$}

\prob{Lexicographically Minimal \SAT with constants  ($\LMS{S}$)}
{An $S$-formula $\Phi$ with constants}
{The lexicographically smallest satisfying assignment of $\Phi$}

\prob{Lexicographically Maximal \SAT  ($\LASN{S}$)}
{An $S$-formula $\Phi$}
{The lexicographically largest satisfying assignment of $\Phi$}

\prob{Lexicographically Maximal \SAT with constants  ($\LAS{S}$)}
{An $S$-formula $\Phi$ with constants}
{The lexicographically largest satisfying assignment of $\Phi$}

%% file: mres.tex
%-*-LaTeX-*-
%
\rcsInfo $Id: mres.tex,v 1.14 1998/03/06 13:02:47 vollmer Exp $
%
%*Main Results
\section{A Dichotomy Theorem for \protect{$\OptP$}}
\label{mres}

There are known algorithms for deciding satisfiability of given formulae in
polynomial time for certain restricted classes of formulae. 
We first observe that these algorithms can easily be modified to
find minimal satisfying assignments. We first consider 
formulae with constants and then turn to the case where no constants are
allowed.

\begin{theorem}
\label{LMSC}
Let $S$ be a  set of logical relations. 
If $S$ is bijunctive, Horn, anti-Horn or affine, 
then we have $\LMS{C} \in \FP$. 
In all other cases $\LMS{C} \not\in \FP$ unless $\P=\NP$.
\end{theorem}

\begin{proof}
For the cases that $S$ is bijunctive, Horn, anti-Horn or affine,
there are well-known polynomial time procedures to {\em decide\/}
satisfiability of a given formula (see e.g.~\cite{pap94}; for the
case of affine $S$ we use Gaussian elimination).

Now we can use the algorithm 
in Fig.~\ref{FindLexMin} for finding the lexicographically
smallest satisfying solution.
This algorithm is an easy modification of an algorithm from
\cite{CrHe97}.
Note that lines 5 and 8 of the algorithm do not
change one of the properties bijunctive, horn, anti-horn and affine; so the
test whether $e$ is satisfiable runs also in deterministic polynomial
time for the modified formula. 
Since we always try first to assign $x_i = 0$ we obtain the
lexicographically smallest satisfying assignment.

Now let $S$ contain at least one relation which is not bijunctive,
one relation which is not Horn, one relation which is not anti-Horn,
and one relation which is not affine.
Then $\LMSN{S}$ cannot be in $\FP$ (unless \P
= \NP), because Proposition \ref{SATDich1} shows that the corresponding 
decision problem (which is the problem of deciding whether there is
{\em any\/} satisfying assignment, not necessarily the minimal one)
is $\log$-complete for $\NP$. \qed
\end{proof}

%\begin{algorithm}
\begin{figure}
\vspace*{4cm}
\normalsize
\begin{algorithmic}[1]
\REQUIRE{Boolean formula $\Phi$ over $S$ with $\var{\Phi} = 
\{x_1,\ldots ,x_n\}$}
\ENSURE{Lexicographically minimal satisfying assignment $A \in \{0,1\}^n$} 
\STATE $e \leftarrow \Phi$;
\IF{($\Phi$ is satisfiable)} 
 \FOR{$i \leftarrow 1$ \textbf{to} $n$} 
  \IF{($e \wedge \neg x_i$ is satisfiable)}
   \STATE $e \leftarrow (e \wedge \neg x_i)$;
   \STATE $A[i] \leftarrow 0$;
  \ELSE
   \STATE $e \leftarrow (e \wedge x_i)$;
   \STATE $A[i] \leftarrow 1$;
  \ENDIF
 \ENDFOR
 \STATE writeln($A$);
\ELSE 
 \STATE writeln(``0'');
\ENDIF 
\end{algorithmic}
\caption{Algorithm to
calculate the lexicographically minimal satisfying assignment}
\label{FindLexMin}
\vspace*{4cm}
\end{figure}
%\end{algorithm}

%The proof is very similar to the proof of Theorem~\ref{LMSN} given below.

\begin{theorem}
\label{LMSN}
Let $S$ be a  set of logical relations. 
If $S$ is \zval, bijunctive, Horn, anti-Horn or affine, 
then we have $\LMSN{S} \in \FP$. 
In all other cases $\LMSN{S} \not\in \FP$ unless $\P = \NP$.
\end{theorem}

\begin{proof}
The case ``\zval'' is obvious. 
For the cases that $S$ is bijunctive, Horn, anti-Horn or affine,
%there are well-known polynomial time procedures to {\em decide\/}
%satisfiability of a given formula (see e.g.~\cite{pap94}; for the
%case of affine $S$ we use Gaussian elimination).
we can use the same algorithms as in the previous theorem
to {\em decide\/} satisfiability, and again
we use the algorithm in Fig.~\ref{FindLexMin} for finding the lexicographically
smallest satisfying solution.
%Now we can use Algorithm~\ref{FindLexMin} for finding the lexicographically
%smallest satisfying solution. Note that lines 5 and 8 of the algorithm do not
%change one of the properties bijunctive, horn, anti-horn and affine; so the
%test whether $e$ is satisfiable runs also in deterministic polynomial
%time for the modified formula. 
%Since we always try first to assign $x_i = 0$ we obtain the
%lexicographically smallest satisfying assignment.

Now let $S$ contain at least one relation which is not \zval,
one relation which is not bijunctive,
one relation which is not Horn, one relation which is not anti-Horn,
and one relation which is not affine.
\begin{description}
\item[Case 1:] There is a relation in $S$ which is not \nval.
Then $\LMSN{S}$ cannot be in $\FP$ (unless \P
= \NP), because Proposition \ref{SATDich2} shows that the corresponding 
decision problem is $\log$-complete for $\NP$.
\item[Case 2:] $S$ is \nval, i.e.~we know that the $0$-vector is not
a satisfying assignment of the given formula but the $1$-vector is;
and we have to solve the question if there is a lexicographically smaller one.
However Proposition \ref{SATSSNPC} shows that the problem of deciding
whether {\em any\/} assignment different from the $0$- or $1$-vector
exists is $\NP$-complete; thus finding the lexicographically 
smallest solution cannot 
be in \FP{} unless $\P=\NP$.
\qed
\end{description}
\end{proof}

Now we know that there are easy (polynomial time solvable) cases
of finding lexicographically minimal satisfying assignments, and 
other cases where under the assumption that $\P\neq\NP$ no efficient
way exists. However this leaves open the possibility that in the latter
case different levels of inefficiency depending on the properties of $S$
can occur. The following two theorems
rule out this possibility. In the case that the lex min sat problem is
not in $\P$ it is already $\MinP$ complete under metric reductions.

We first consider the (easier) case of formulae where constants are
allowed.

\begin{theorem}
\label{ClassOptP}
Let $S$ be a  set of logical relations. 
If $S$ does not fulfill the properties Horn, anti-Horn, bijunctive or
affine then \LMS{S} is \redm-complete for \MinP. 
\end{theorem}

\begin{proof}
Obviously $\LMS{C} \in \MinP$. Now we have to proof \redm-hardness
for \MinP.

If $S$ does not fulfills the properties Horn, anti-Horn, bijunctive or
affine then Proposition \ref{RepIsAll} shows that $\rep{S}$ includes all
boolean relations.

Let $R_i$ be any logical relation. Proposition \ref{RepIsAll}
tells us that there exists an $S$-formula 
$\Phi = \exists y_1 \dots \exists y_k \Phi'$, 
representing $R_i$, where $\Phi'$ contains no quantifier. Any clause of
a \TSAT formula can be represented by a finite number of boolean
relations. So any clause $C_i$ of a \TSAT formula $\Phi$ can be represented 
by an $S$-formula $\Phi_i$. $\var{\Phi_i}$ consists of the variables in
$\var{C_i}$ plus a number of variables of the form $y_j$. We pick different
sets of $y_j$-variables for different formulae $\Phi_i$.

Now we construct a function $g_2 \in \FP$ mapping a \TSAT formula $\Phi$ 
into an $S$-formula $\Phi'$ by replacing each $C_i$ by the corresponding 
$\Phi_i'$, where $\var{\Phi'}$ consists out of $\{\enu{x}{1}{n}\}$
plus a set of variables of the form $y_j$.
We order the variables by their index and by alphabet, 
i.e.~$x_1<x_2<x_3<\cdots<y_1<y_2<\cdots$.
Note that we can drop the $\exists$-quantifiers of
the variables $y_j$ since we ask for a satisfying assignment of $\Phi'$.
The ordering of the variables ensures that in the minimal satisfying 
assignment of $\Phi'$ the variables in $\{\enu{x}{1}{n}\}$ will be minimal
with respect to satisfaction of $\Phi$.

Now the function $g_1 \in \FP$ shortens the assignment and removes all
bits belonging to the variables $y_j$. Thus 
$g_1$ applied to the minimal satisfying assignment of $\Phi'=g_2(\Phi)$
produces the minimal satisfying assignment for $\Phi$. 
This says that $\LMT \redm \LMS{C}$.
\qed
\end{proof}

Note that our proof heavily hinges on Schaefer's Proposition~\ref{RepIsAll}.
However Schaefer's technique always introduces new variables, which pose no
problem in his context, but are not allowed here. We can only
remove these new variables in the end because we have the power of 
metric reductions.

Mainly we are interested in formulae without constants. So we have
to get rid of the constants in the construction of the just given
proof. This is achieved in the reduction which we now present.

\begin{theorem}
\label{CompleteNoConstants}
Let $S$ be a  set of logical relations. 
If $S$ is not \zval, Horn, anti-Horn, bijunctive or affine, then \LMSN{S}
is \redm-complete for \MinP.
\end{theorem}

\begin{proof}
Clearly $\LMSN{S}\in\MinP$.
We want to show that $\LMS{S}$ reduces to $\LMSN{S}$.
\begin{description}
\item[Case 1:] $S$ is not \nval.

Using Proposition~\ref{StructOfRepS} we know, that 
$[ x ], [ \neg x ] \in \repn{S}$ 
or $[ x \not\equiv y ] \in \repn{S}$. In what follows, we again sort all 
variables by index and alphabet.
\begin{description}
\item[Case 1.1:] $[ x ], [ \neg x ] \in \repn{S}$.

Let $\Phi$ an $S$-formula
with constants and $\var{\Phi} = \{ \enu{x}{1}{n} \}$. Now we can
remove the constants by replacing any $1$ by $y_1$ and $0$
by $y_0$ and adding clauses representing $\{y_1\}$ and $\{\neg y_0\}$. 
Define the function $g_2$ such that $g_2(\Phi)$ performs exactly
the just described replacement.

Now $I \minmodels \Phi$ if and only if 
$I' \eqd (I \cup \{ y_0:=0, y_1:=1 \}) \minmodels \Phi'$, 
where $\Phi'\eqd g_2(\Phi)$. The function $g_1$ removes the last two bits 
(assignments of $y_0$ and $y_1$) from $I'$, showing that 
$\LMS{C} \redm \LMSN{S}$.
\item[Case 1.2:] $[ x \not\equiv y ] \in \repn{S}$. 

Let $\Phi$ an
$S$-formula with constants and $\var{\Phi} = \{ \enu{x}{1}{n} \}$.
We construct an $S$-formula 
$\Phi' \eqd \subst{\subst{\Phi}{0}{u}}{1}{v} \wedge (u \not\equiv v)$ 
without constants. 
Define $g_2$ by $g_2(\Phi)=\Phi'$.
Now suppose there exists a satisfying assignment 
$I' \eqd I_{w} \cup \{ u := 1, v := 0 \}$. %  where $I_{w} \minmodels\Phi$. 
This would be an unwanted assignment, since $v$ should represent
$1$ and $u$ should represent $0$. But there exists also the correct
satisfying assignment $I'' \eqd I_{r} \cup \{ u:=0, v:=1 \}$,
where $I_{r} \minmodels\Phi$. 
This assignment is clearly lexicographically
smaller than $I'$ and thus $I'' \minmodels \Phi'$ iff 
$I_{r} \minmodels \Phi$.
  
Now we remove the assignment for $u$ and $v$ by $g_1$. The functions $g_1$ and
$g_2$ show that 
$\LMS{S} \redm \LMSN{S}$.
\end{description}
\item[Case 2:] $S$ is \nval.

Having an $S$-formula with constants we construct one without constants
in polynomial time by $g_2$ as follows. Let $R \in S$ a relation which
is not \zval but \nval and $\Phi' \eqd \subst{\subst{\Phi}{0}{u}}{1}{v} \wedge 
R(v, \ldots ,v)$. We claim that $I \minmodels \Phi$ \gdw $I \cup \{ u := 0, 
v := 1 \} \minmodels \Phi'$.

First suppose that $I \minmodels \Phi$. It is clear from the clause
$R(v, \ldots ,v)$ that we have to choose $v:=1$. Since we are
interested in the lexicographically smallest solution we have to choose $u:=0$
giving us immediately $I \cup \{ u := 0, v := 1 \} \models \Phi'$ and 
certainly also $I \cup \{ u := 0, v := 1 \} \minmodels \Phi'$.
Now let $I \cup \{ u := 0, v := 1 \} \minmodels \Phi'$. Suppose that
there exists a satisfying solution $I_s$ for $\Phi$ being lexicographically
smaller than $I$. Obviously $I_s \cup \{ u := 0, v := 1 \}$ is a
lexicographically smaller satisfying assignment than $I \cup \{ u := 0, v := 1
\}$ giving us a contradiction to $I \cup \{ u := 0, v := 1 \} 
\minmodels \Phi'$.

We remove the assignment for $u$ and $v$ by $g_1$, showing that 
$\LMS{C} \redm \LMSN{S}$. 
\qed
\end{description}
\end{proof}

Observe that Schaefer's Proposition~\ref{StructOfRepS} is not
sufficient to obtain the above result. Our proof substantially depends on the
ability to force a suitable ordering of the assignments by ordering the
variables in a reasonable way.

Thus we get dichotomy theorems for finding lexicographically minimal
satisfying assignments of propositional formulae, both for the case
of formulae with constants and without constants.

\begin{corollary}[Dichotomy Theorem for \lms with constants]
\label{DichMinPC}
Let $S$ be a  set of logical relations. 
If $S$ is bijunctive, Horn, anti-Horn or affine, then we have
$\LMS{C} \in \FP$. In all other cases $\LMS{C}$ is $\redm$-complete 
for $\MinP$.
\end{corollary}

\begin{corollary}[Dichotomy Theorem for \lms]
\label{DichMinPN}
Let $S$ be a  set of logical relations. 
If $S$ is \zval, bijunctive, Horn, anti-Horn or affine, then we have
$\LMSN{S} \in \FP$. In all other cases $\LMSN{S}$ is $\redm$-complete 
for $\MinP$.
\end{corollary}

If we compare the classes of relations in the statements of the above
corollaries with those needed in Schaefer's results 
(Propositions~\ref{SATDich1} and \ref{SATDich2}), the following 
consequence is immediate:

\begin{corollary}
Let $S$ be a  set of logical relations.
\begin{enumerate}
\item $\SATS{S}$ is $\NP$-complete if and only if 
$\LMS{S}$ is $\MinP$ complete.
\item If $\SATSN{S}$ is $\NP$-complete then $\LMSN{S}$ is $\MinP$ complete.
\item If $S$ is a set of logical relations which is \nval but is not 
\zval, Horn, anti-Horn, bijunctive, or affine, then $\SATSN{S}$ is
in $\P$ but $\LMSN{S}$ is $\MinP$ complete.
\end{enumerate}
\end{corollary}

The above corollary completely clarifies the connection between
decision and optimization for the optimal assignments problem.

\begin{example}
Hierarchical \SAT is the variant of \TSAT where only unnegated variables
occur and we require that in each clause if either the first or the second
variable are satisfied then the third variable is not satisfied, and
if the third variable is satisfied then also the first and second variable
are satisfied. In our framework this problem is given by
$S=\{R\}$, where $R=\{(1,0,0),(0,1,0),(1,1,1)\}$. It can
be seen using techniques from \cite{Sch87} that $S$ is
\nval but is not \zval, Horn, anti-Horn, bijunctive, or affine.
Thus $\SATSN{S}$ is in $\P$ but $\LMSN{S}$ is $\MinP$ complete.
\end{example}

%Results analogous to the above for the problem of finding maximal
%assignments can be proved, where we just have to replace \nval by
%\zval.

Results analogous to the above for the problem of finding maximal
assignments can be proved:

\begin{theorem}[Dichotomy Theorem for \las] 
Let $S$ be a  set of logical relations.
\begin{enumerate}
\item If $S$ is bijunctive, Horn, anti-Horn or affine, then 
$\LAS{C} \in \FP$. Otherwise $\LAS{C}$ is $\redm$-complete for $\MaxP$.
\item If $S$ is \nval, bijunctive, Horn, anti-Horn or affine, then 
$\LASN{S} \in \FP$. Otherwise $\LASN{S}$ is $\redm$-complete for $\MaxP$.
\end{enumerate}
\end{theorem}

If we look at the definition of metric reductions (see Sect.~\ref{easy})
and compare this with the proofs given above, we see that we do not
need the full power of metric reductions here. In fact the function
$g_1$ in our proof is a function which, first, does not depend on
$x$ but only on $g_2(x)$, and second, $g_1$ is ``almost'' the identity
function---$g_1(z)$ is obtained from $z$ by simply stripping away a few
bits. Since $g_1$ is almost the identity, let us call these
reductions \emph{weak many-one reductions;} that is, $f$ is weakly
many-one reducible to $h$ if there are two functions $g_1,g_2\in\FP$
where $g_1(z)$ is always a sub-word of $z$, such that for all $x$,
$$f(x)=g_1(h(g_2(x))).$$

\begin{theorem}\label{weakred}
All the above given completeness results also hold for
weak many-one reductions instead of metric reductions.
\end{theorem}

\begin{proof}
A close look at Krentel's work shows that Proposition \ref{krentelcomplete}
also holds for weak many-one reductions. The reductions given above in
the proofs of Theorems~\ref{ClassOptP} and \ref{CompleteNoConstants}
are in fact weak many-one reductions. Since these reductions are transitive
our theorem follows.
\qed
\end{proof}

The question that now arises is of course if we can even prove our
completeness results for many-one reductions, which are 
weak many-one reductions 
where $g_1$ is the identity function. However this cannot be expected for
``syntactic'' reasons, since when we manipulate a given formula $\Phi$
constructing $\Phi'$ such that $\var{\Phi}\neq\var{\Phi'}$
then an assignment of $\Phi'$ simply by definition
cannot be an assignment of $\Phi$. And it seems that there is no way 
of getting around this; we have to change the variable set.
%On a more formal level we can show that if Corollary~\ref{DichMinPN} holds for
%many-one reductions then $\P=\NP$. 

%% file: fpnp.tex
%-*-LaTeX-*-
%
\rcsInfo $Id: fpnp.tex,v 1.9 1998/03/06 13:02:43 vollmer Exp $
%
%*FP^NP
\section{A Dichotomy Theorem for \protect{$\P^{\NP}$}}
\label{fpnp}

Given a function $f\colon\N\rightarrow\N$, define
the set $L_f = \{x\in\Sigma^*\mid f(x) \equiv 1\pmod{2}\}$.
Often it turns out that if $f$ is complete for $\OptP$ under 
metric reductions, then the set $L_f$ is complete for $\P^{\NP}$ under usual 
many-one reductions; a precise statement is given below.

In our context the above problem translates to the question if
the largest variable in a lexicographically minimal assignment 
of a given $S$-formula gets the value $1$. Let us denote this problem
by $\OMASN{S}$, and in the case that $S$-formulae with constants are
allowed by $\OMAS{S}$. (In the case of maximal assignments we use the notation
$\OAAS{S}$ and $\OAASN{S}$.)
The corresponding problems for unrestricted propositional formulae will be 
denoted by $\omas$ and $\oaas$.

\begin{proposition}[\cite{kre88}]
$\omas$ and $\oaas$ are complete for the class
$\P^{\NP}$ under many-one reductions.
\end{proposition}

It is known that if $f$ is complete for $\MinP$ or $\MaxP$
under many-one reductions (see the discussion at the end of
Sect.~\ref{mres}) then 
$L_f$ is complete for $\P^{\NP}$ under usual many-one reductions 
\cite{kre88}, see also \cite{vol94b}.
In the case that $f$ is only metric complete or weakly many-one complete, 
a similar result is not known.
Since in Sect.~\ref{mres} we proved completeness under 
weak many-one reductions
we cannot by the above remark mechanically translate our results for
$\SATSN{S}$ to completeness results for $\OMASN{S}$ for the class
$\P^{\NP}$. However by separate proofs we can determine the complexity
of $\OMAS{S}$ and $\OMASN{S}$.

\begin{theorem}[Dichotomy Theorem for \omas with constants]
Let $S$ be a set of logical relations. 
If $S$ is bijunctive, Horn, anti-Horn or affine, 
then we have $\OMAS{S} \in \P$. 
In all other cases $\OMAS{S}$ is complete for $\P^{\NP}$ under
many-one reductions.
\end{theorem}

\begin{proof}
If $S$ is bijunctive, Horn, anti-Horn or affine, 
then $\OMAS{S} \in \P$, since we can use Algorithm~\ref{FindLexMin}
to find the minimal assignment, and then we accept if and only if
the truth value $1$ is assigned to the largest variable.

In the other cases we reduce $\omats$ to $\OMAS{S}$. 
In the proof of Theorem~\ref{ClassOptP} we showed how to transform
an arbitrary formula $\Phi$ with $\var{\Phi}=\{\enu{x}{1}{n}\}$ into an 
$S$-formula at the cost of introducing new variables of the form $y_j$.
We modify this construction as follows: Introduce one more variable $z$
(larger than all the other variables). Transform $\Phi$ into
$\Phi'$ as described in Theorem~\ref{ClassOptP}. Finally set
$\Phi''=\Phi'\land(x_n\equiv z)$. (Observe that the predicate $\equiv$ 
is in $\rep{S}$.)
Let $I,I',I''$ be the minimal satisfying assignments of $\Phi$, $\Phi'$
and $\Phi''$. Observe that they all agree on assignments of the variables
in $\var{\Phi}$. Now we have
$$I(x_n)=I'(x_n)=I''(x_n)=I''(z).$$
Thus $\Phi\in\omats$ if and only if $\Phi''\in\OMAS{S}$, which proves
the claimed hardness result.
\qed
\end{proof}

\begin{theorem}[Dichotomy Theorem for \omas]
Let $S$ be a set of logical relations. 
If $S$ is \zval, bijunctive, Horn, anti-Horn or affine, 
then we have $\OMASN{S} \in \P$. 
In all other cases $\OMASN{S}$ is complete for $\P^{\NP}$ under
many-one reductions.
\end{theorem}

\begin{proof}
Similar to the proof of the previous theorem. The easy case is obvious.
In the hard case define $\Phi''$ as above, and then use
the construction of Theorem \ref{CompleteNoConstants} to remove the constants.
Let $\Phi'''$ be the resulting formula.
The variables introduced in this last step should be smaller than $z$.
Then we can argue as in the previous proof that $z$ is assigned one
in a minimal assignment for $\Phi'''$
if and only if $x_n$ is assigned one in a minimal assignment for $\Phi$. 
\qed
\end{proof}

Again, analogous results for maximal assignments can be proved:

\begin{theorem}[Dichotomy Theorem for \oaas]
Let $S$ be a set of logical relations.
\begin{enumerate}
\item If $S$ is bijunctive, Horn, anti-Horn or affine, 
then $\OAAS{S} \in \P$. 
In all other cases $\OAAS{S}$ is complete for $\P^{\NP}$ under
many-one reductions.
\item If $S$ is \zval, bijunctive, Horn, anti-Horn or affine, 
then $\OAASN{S} \in \P$. 
In all other cases $\OAASN{S}$ is complete for $\P^{\NP}$ under
many-one reductions.
\end{enumerate}
\end{theorem}

%% file: concl.tex
%-*-LaTeX-*-
%
\rcsInfo $Id: concl.tex,v 1.2 1998/02/23 11:17:07 vollmer Exp $
%
%*Conclusion
%\section{Conclusion}

%The presented results completely clarify the complexity of 
%optimization problems based on lexicographic ordering of
%satisfying assignments of propositional formulae.

\medskip

\noindent{\bf Acknowledgment.} We are extremely grateful to
Nadia Creignou, Caen, for a lot of helpful hints.